\documentclass[structabstract]{aa}  
\usepackage{graphicx}
\usepackage{txfonts}
\usepackage{natbib}
\bibpunct{(}{)}{;}{a}{}{,} 

\begin{document}

   \title{An active region filament studied simultaneously in the chromosphere and photosphere: I -- Magnetic structure}


   \author{C. Kuckein\inst{1,2}
    \and V. Mart\'\i nez Pillet\inst{1}
    \and R. Centeno\inst{3}}   
  
   \institute{Instituto de Astrof\'\i sica de Canarias, V\'\i a 
             L\'{a}ctea s/n, E-38205 La Laguna, Tenerife, Spain\\
             \email{ckuckein@iac.es}
   \and Departamento de Astrof\'\i sica, Universidad de La Laguna, E-38206 La Laguna, Tenerife, Spain
   \and High Altitude Observatory (NCAR), Boulder, CO 80301, USA}

  \date{Received date / Accepted date}

 
  \abstract
   {}
   {A thorough multiwavelength, multiheight study of the vector magnetic
   field in a compact active region filament (NOAA 10781) on 2005 July 3rd and
   5th is presented. We suggest an evolutionary scenario for this filament. }
   {Two different inversion codes were used to analyze the full Stokes vectors
   acquired with the Tenerife Infrared Polarimeter (TIP-II) in a spectral range
   which comprises the chromospheric \ion{He}{i} 10830\,\AA\ multiplet and the
   photospheric \ion{Si}{i} 10827\,\AA\ line. In addition, we used
   \textit{SOHO}/MDI magnetograms as well as BBSO and
   \textit{TRACE} images to study the evolution of the filament and its active
   region (AR). High resolution images of the Dutch Open Telescope were also
   used. }
   {An active region filament (that was formed before our observing run) was
   detected in the chromospheric Helium absorption images on July 3rd.  The
   chromospheric vector magnetic field in this portion of the filament
   was strongly sheared (parallel to the filament axis) whereas the
   photospheric field lines underneath had an inverse polarity configuration.
   From July 3rd to July 5th, an opening and closing of the polarities at
   either side of the polarity inversion line (PIL) was recorded, resembling 
   the recently discovered process of the sliding door effect seen by Hinode. 
   This is confirmed both with TIP-II and \textit{SOHO}/MDI data. During this time,
   a newly created region that contained pores and orphan penumbrae at the 
   PIL was observed. On July 5th, a normal polarity configuration was inferred
   from the chromospheric spectra, while strongly sheared field lines aligned with
   the PIL were found in the photosphere. In this same data set, the spine of the 
   filament is also observed in a different portion of the FOV and is clearly
   mapped by the Silicon line core.
   }
   {The inferred vector magnetic fields of the filament suggest a flux rope
   topology. Furthermore, the observations indicate that the filament is
   divided in two parts, one which lies in the chromosphere and another one
   that stays trapped in the photosphere. Therefore, only the top of the helical
   structure is seen by the Helium lines. The pores and orphan penumbrae at the PIL
   appear to be the photospheric counterpart of the extremely low-lying filament.
   We suggest that orphan penumbrae are formed in very narrow PILs of compact ARs 
   and are the photospheric manifestation of flux ropes in the photosphere.}
   \keywords{Sun: filaments, prominences -- 
		Sun: photosphere --
		Sun: chromosphere --
		Sun: magnetic topology --
                Techniques: polarimetric
               }

   \authorrunning{Kuckein et al.}
   \titlerunning{Simultaneous study of the vector magnetic field in an AR filament}
   \maketitle

\section{Introduction}
Solar filaments, also called prominences when observed in emission outside the
disk, are elongated structures formed of dense plasma which lies above polarity
inversion lines (PILs) of the photospheric magnetic field \citep{babcock55} and
has a lower temperature than its surroundings. A functional definition of the PIL
would be an imaginary line that separates two close areas of opposite
polarity. On disk filaments are readily identifiable in the quiet Sun 
(quiescent filaments) and in active regions (active region filaments) when observed
using common chromospheric wavelengths, e.g., the \ion{He}{i} 10830\,\AA\ multiplet
and the H$\alpha$ 6563\,\AA\ line. The magnetic field plays a major role in their
formation, stability and evolution. Hence, spectropolarimetric observations at
different heights of the atmosphere combined with high angular resolution
images are needed to obtain an overall picture of the physical processes which take
place in filaments. The magnetic field strength and its orientation can be inferred
using appropiate diagnostic techniques that are able to interpret the Zeeman effect
as well as scattering polarization and its modification through the Hanle effect 
\citep[][and references
therein]{tandberg95}. Although many line-of-sight observations have been carried out
in the past decades, it is important to emphasize the need for full-Stokes measurements in
order to have complete information of the vector magnetic field. 

According to observational studies, the magnetic field topology under active region (AR)
filaments has sheared or twisted field lines along the PIL which can support
dense plasma in magnetic dips \citep{lites05, lopez06}. Previous 
findings about prominence magnetic structure have shown models with dipped field
lines in a normal \citep{kippen57} or inverse\footnote{Throughout this paper,  
by {\em inverse configuration} we mean a
magnetic field vector, perpendicular to the filament's axis, that
points from the negative to the positive polarity. This is the opposite to what
one would expect in a potential field solution - where the field points from the
positive to the negative side (and is referred to as normal
configuration).} \citep{kuperus74, pneuman83}
polarity configuration, the latter having a helical structure.
From a large sample of quiescent prominence observations,
\citet{leroy84} found both types of configurations and a strong correlation
between the magnetic field topology and 
the height of the prominence. On the other hand, \citet{bommier94} found predominantly
inverse polarity configurations. However, more recent photospheric
spectropolarimetric observations of AR filaments have revealed that both
configurations can coexist in the same filament \citep{guo10} or even evolve
with time from one type to the other, as presented by \citet{okamoto08} from
the analysis of vector magnetogram sequences.

The formation process of filaments is still a controversial issue in solar
physics and has been widely debated. On the one hand, there is the \emph{sheared arcade model}
which, by large scale photospheric footpoint motions (such as shear at, 
and convergence towards the PIL, and subsequent reconnection processes) 
is able to form dips, and even helical structures, where plasma can be
gravitationally confined \citep[e.g.,][]{pneuman83,balle89,
antiochos94,aulanier98, devore00,martens01,welsch05,karpen07}. 
This model is capable of reproducing both inverse and normal polarity
configurations in the same filament \citep{aulanier02}. However, recent
observational works in a quiescent filament \citep{hindman06} and along an 
AR filament channel \citep{lites10} did not find evidence for these
systematic photospheric flows that converged at the PIL and 
triggered reconnection processes. In our opinion, more observations are needed
to support this important result. 

On the other hand, the \emph{flux rope emergence model}
assumes that the twist in the field lines is already present in the 
convection zone before emerging 
into the atmosphere \citep[eg.,][]{kurokawa87, low94, low95, lites05}. The rise
of twisted magnetic fields 
has been studied by various authors through observations
\citep{lites95, leka96, lites10} as well as three-dimensional (3D) numerical
simulations \citep[e.g.,][]{fan01,archontis04,magara04,sykora08,fan09,yelles09}. Such an emerging
helical flux rope scenario, combined with the presence of granular flows, 
is suggested to be the
main process of formation and maintenance of the AR filament studied by
\citet{okamoto08, okamoto09}. Recent nonlinear force-free field (NLFFF)
extrapolations of photospheric magnetic fields underneath AR filaments
\citep{guo10, canou10, jing10} agreed that the main structure 
has to be a flux rope, although \citet{guo10} also found dipped
arcade field lines in the same filament. 
Flux rope emergence from below the photosphere encounters severe
difficulties to reach chromospheric layers and above, as it is loaded
with mass \citep[e.g.,][]{archontis08}. Indeed, the previous work suggests
that a second flux rope formed from instabilities and atmospheric
reconnection is what lifts up the mass and forms the observed active region
filaments. 

The main difference between the sheared arcade and the emerging flux rope 
models is the existence of a flux concentration stuck at photospheric levels in the 
latter scenario. The reader is referred to the paper of
\citet{mackay10} for a recent review on the magnetic structure of filaments.

In the past, only a few measurements of the magnetic field strength in AR
filaments have been done. For instance, \citet{kuckein09}, using the same data set that 
is described in this paper, found a predominance of Zeeman-like signatures in the 
Stokes profiles. Using three different methods, they inferred very strong magnetic fields
in the filament (600--700\,Gauss), with a dominant transverse component (with respect to 
the line-of-sight). 

The aim of this paper is to study the strength and topology of the magnetic field
in an active region filament at photospheric and chromospheric heights 
simultaneously. Recently, several studies have presented analyses of the vector 
magnetic field in filaments or prominences from observations either in the photosphere
\citep[e.g.,][]{lites05,lopez06,okamoto08,guo10,lites10} or the chromosphere
\citep[e.g.,][]{lin98,casini03,merenda06,kuckein09}, but non of them have
inferred the field at both heights at the same time. The 
10830\,\AA\ spectral region, which includes a chromospheric \ion{He}{i} triplet and a
photospheric \ion{Si}{i} line, offers a unique spectral window to understand the physical
processes that take place in AR filaments as already shown by 
\citet{sasso11}. In this work, we focus on the overall magnetic configuration 
observed simultaneously in the photosphere and the chromosphere.

\section{Observations}
The studied active region filament, NOAA AR 10781, was observed on 2005 July 3rd and
5th using the Tenerife Infrared Polarimeter \citep[TIP-II,][]{tip2}
at the German Vacuum Tower Telescope (VTT, Tenerife, Spain). 
TIP-II acquires images at 4 different modulation states and
combines them in order to measure the Stokes parameters ($I$, $Q$, $U$ and $V$)
along the spectrograph slit. 

\begin{figure}
 \resizebox{\hsize}{!}{\includegraphics{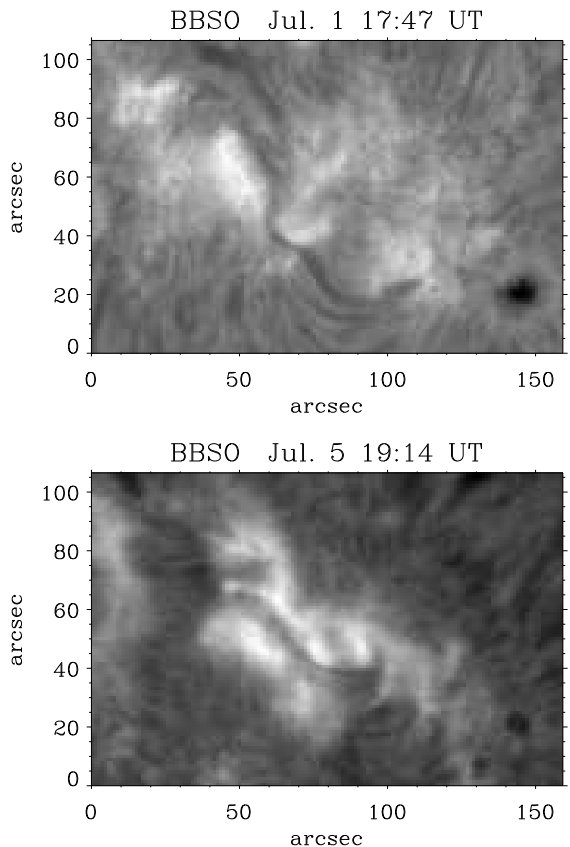}}
  \caption{Big Bear H$\alpha$ image of the active region under study. The
  leader sunspot is seen in the bottom right corner. The filament is seen on 
  both days, July 1st and 5th. Note the more compact nature of the plage
  emission on the latter day.}
 \label{Fig:BBfil}
\end{figure}

The latitude and longitude of the filament region was N16-E8 (around $\mu \sim 0.95$) for 
the first day, and N16-W18 ($\mu \sim 0.91$), for the second. During the observations, 
real time H$\alpha$ images and \textit{SOHO}/MDI \citep{mdi95} magnetograms were used as 
a reference to position the slit (0\farcs5 wide and 35\arcsec long) at the center of the AR, on top of
the polarity inversion line. Figure \ref{Fig:BBfil} shows two Big Bear H$\alpha$ images
from July 1st and 5th where the filament can easily be recognized. On July 1st, 
the filament shows a more diffusive nature than on July 5th, when it
displays a rather compact configuration with brighter H$\alpha$ emission
from the plage flanking it. Scans were taken with TIP-II from east to west with the slit
parallel to the PIL and step sizes of 0\farcs4 for July 3rd and 0\farcs3 for
July 5th,  making up at least a field of view (FOV) of $\sim 30\arcsec \times 35\arcsec$, with a pixel 
size along the slit of 0\farcs17. 

The TIP-II spectral range spanned from $10825$ to
$10836$\,\AA\ with an original spectral sampling of \mbox{$\sim 11.04$ m\AA\, px$^{-1}$}. 
This spectral window included the photospheric \ion{Si}{i} line at 10827\,\AA\ and the chromospheric
\ion{He}{i} triplet at 10830\,\AA. It also contained at least one telluric 
H$_2$O line which is used for the absolute velocity calibration.

All images were corrected for flat field, dark current and calibrated 
polarimetrically following standard procedures
\citep{collados99,collados03}. The adaptive optics system of the
VTT \citep[KAOS,][]{kaos} was locked on nearby pores and orphan penumbrae, i.e., penumbral-like
structures not connected to any umbra \citep[a term coined by][]{zirin91}, and highly improved the 
observations which had changeable seeing
conditions. A binning of 3 pixels in the spectral domain, 6\,px
along the slit and 3\,px along the scanning direction was carried out to improve
the signal-to-noise ratio needed for the full Stokes spectral line inversions. 
Thus, the final spectral sampling was $\sim 33.1$\,m\AA\,px$^{-1}$. The 
spatial resolution, when the KAOS
system was locked, reached $\sim 1$\arcsec. However, the binned data 
used for the inversions (except when stated otherwise) had a
resolution of 2\arcsec.

The \ion{He}{i} triplet comprises three spectral lines which, according to the
\textit{National Institute of Standards and Technology} (NIST), are the
10829.09\,\AA\ ``blue'' component and the ``red'' components at
10830.25\,\AA\ and 10830.34\,\AA\ lines. The latter two are blended and
consequently appear as a single spectral line in the solar spectrum. The
formation height of the Helium triplet happens in the upper chromosphere 
\citep{avrett94} and therefore
it is especially interesting for the study of filaments and their magnetic properties,
as already proved by many authors
\citep{lin98,trujillo02,merenda07,casini09,kuckein09,sasso11}. The strong
photospheric \ion{Si}{i} absorption line at 10827.089\,\AA\ originates between
the terms $^3\mathrm{P}_2 \rightarrow\, ^3\mathrm{P}_2$ and has a Land\'e
factor of $g_\mathrm{eff} = 1.5$. The combination of these lines is an
excellent diagnostic tool to study magnetic fields and their coupling between
the photosphere and the chromosphere. 

   \begin{figure}
   \resizebox{\hsize}{!}{\includegraphics{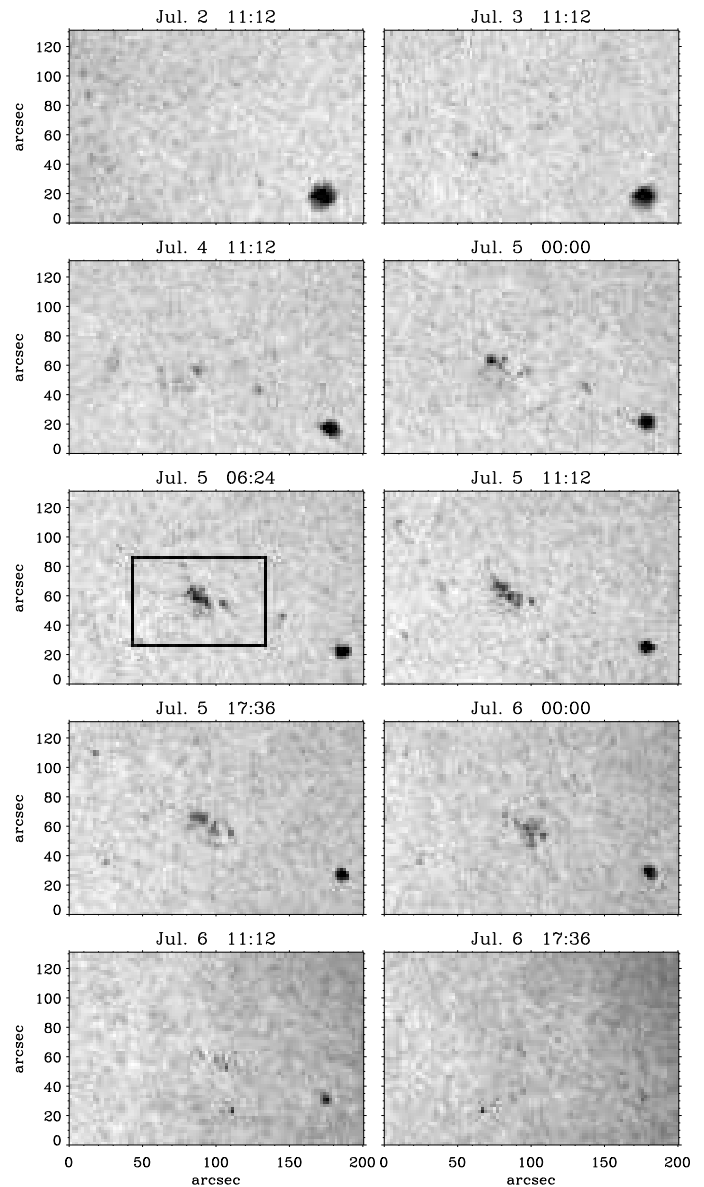}}
   \caption{Time sequence of \textit{SOHO}/MDI continuum images of the whole
   active region NOAA 10781 between July 2nd and 6th of 2005. The black box
   shows the size and location of the maps of Fig. \ref{Fig:MDIevol} which have
   a smaller but more detailed FOV. A developed sunspot can be seen at the
   \textit{bottom right} corner of the AR. Between July 2nd and 3rd the center
   of the AR was almost devoid of pores. On July 5th we increased the cadence
   of the sequence in order to show the quick appearance of new pores and
   orphan penumbrae at the PIL during that day. }
   \label{Fig:MDIcont}
    \end{figure}

  \begin{figure}
   \resizebox{\hsize}{!}{\includegraphics{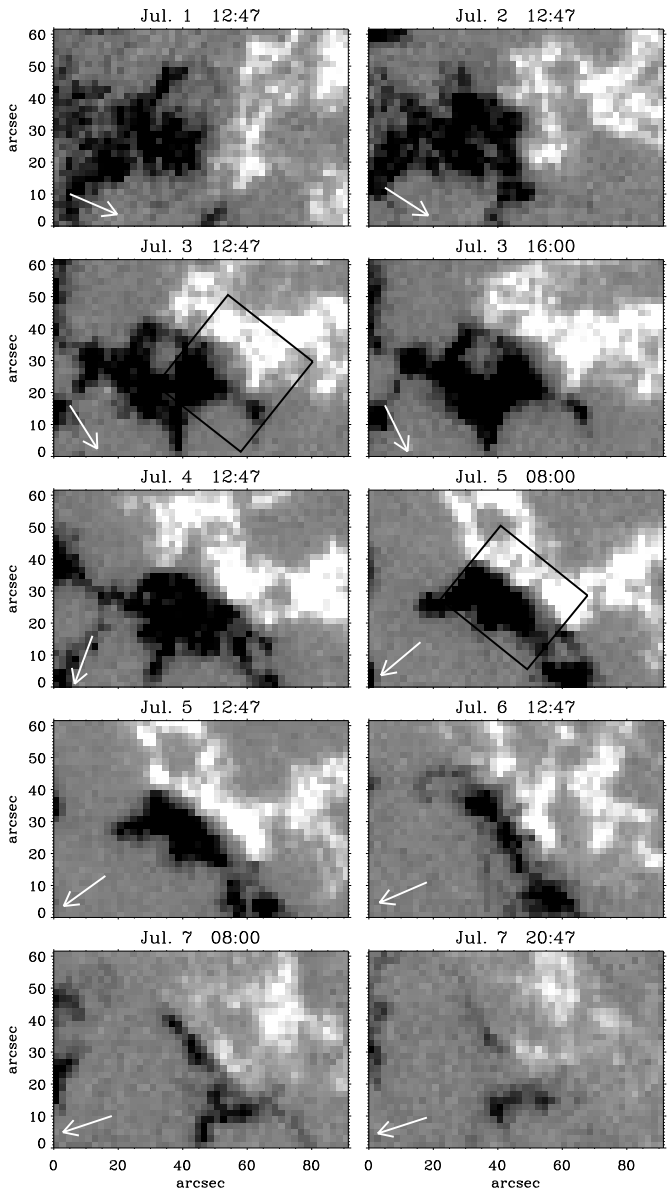}}
   \caption{\textit{SOHO}/MDI line-of-sight magnetogram evolution of the
   compact plage between July 1st and 7th of 2005. Note that the date and time
   of each panel does not correspond to the panels in Fig. \ref{Fig:MDIcont}.
   Images are saturated at $\pm400$\,G. Solar north and west correspond to up
   and right respectively. The black boxes show the approximate scanned area 
   of TIP-II for
   both days. On July 3rd, the space between both polarities broadens in the
   lower part of the PIL, whereas in the following days the whole AR is more
   compact. The size of the panels corresponds to the black rectangle in Fig.
   \ref{Fig:MDIcont}. White arrows indicate the direction to disk center.} \label{Fig:MDIevol}
    \end{figure}

In this paper we also report on \textit{SOHO}/MDI images which are used to understand
the long term (days) evolution of the AR in the PIL region.

\subsection{Evolution of the AR (SOHO/MDI)} \label{sec:MDIevol}
Active region NOAA 10781 emerged some weeks before our observing run on
the back side of the Sun. Its magnetic configuration, as seen by \textit{SOHO}/MDI
(see below), clearly corresponds to that of an AR that is in its decay phase,
i.e., a round leader sunspot followed by facular regions of both
polarities that show the latitudinal shear produced by the action of
surface flows \citep[e.g.,][]{vanballe08}.
The filament is found above the PIL in the plage region.
The time sequence of continuum images of MDI between 2005 July
2nd and 6th is presented in Fig. \ref{Fig:MDIcont}. Since 
filaments are not visible here,
we expanded the FOV in order to use the AR leader 
sunspot, located in the lower right corner, as a reference.
As shown in Fig. \ref{Fig:MDIcont}, it is not until July 4th that small pores
start to emerge and gather together. White light images of the 
\textit{Transition Region
and Coronal Explorer} (\textit{TRACE}) confirm this behavior. On
July 5th, we increased the cadence in Fig. \ref{Fig:MDIcont}
to show that more pores quickly emerged and orphan penumbrae formed
(see black rectangle on the July 5th 06:24 UT panel). 
Interestingly, these orphan penumbral regions
act like a bridge connecting different small groups of pores
together.  However, during July 6th the pores and orphan penumbrae 
disappear completely. It is also worth noting that the leading
sunspot of the AR, which on July 2nd had a round and rather symmetric shape,
also decays away slowly over this period of time and almost vanishes by the
end of July 6th. 

Figure \ref{Fig:MDIevol} provides line-of-sight (LOS) magnetograms from MDI starting
on 2005 July 1st until July 7th. This period was chosen in order to carry out a 
detailed study of the
magnetic morphology and evolution of the AR. Note that the FOV is
smaller and the panels show different dates and times than Fig.
\ref{Fig:MDIcont}. The AR and its PIL are well defined. 
The \textit{top
left} image shows the two polarity regions as of July 1st. 
On the second day, the AR became
more compact and the gray area in between the two opposite polarities, i.e. the PIL,
developed a winding shape that pointed approximately in the N-S direction. 
The next two images (from July 3rd) show the AR becoming even more compact, since the
black and white polarities approached each other. It is important to stress that,
on this day, the AR rotated counterclockwise and the PIL oriented itself at 
$\sim 45^\circ$ with respect to solar north. These images
also reveal that from July 2nd to 3rd, although the whole AR became more
compact, the lower-right part of the PIL first broadened then become narrower again on
the following day (a process that is also seen on the upper-left part of the PIL
one day later).
By July 5th, the PIL was highly compact with the two polarities in close
contact in the scaling of the figure ($\pm$ 400G).
This behavior is consistent with the
observations of \citet{okamoto08} who described the opening and closing of the
PIL region under an AR filament as a ``sliding door''. The authors
found observational evidence for the emergence of a flux rope under that AR
filament. 

\subsection{Magnetic flux evolution of the AR}

Recently, the study of AR filaments has been put in the context of the global
flux history of the active region as a whole \citep[][]{vanballe07}. The
seminal work of \citet{balle89} showed that flux ropes (and filaments) can be
formed through successive cancellations at the PIL, transforming active region
flux into flux rope field lines. This process submerges as much flux below the surface
as is injected into the flux rope itself. Thus, evaluating the flux
losses that occur during the last stages of an active region is important to
understand this evolutionary phase. Measurements of such flux losses in an active region by 
\citet{sterling10} yielded $2.8 \times 10^{21}$\,Mx day$^{-1}$.
More recently, for a medium sized active region, \citet{green11} quoted
a smaller rate of $2.5 \times 10^{20}$\,Mx day$^{-1}$. This number agrees better
with the estimate by \citet{sainz08}, who found $4.8 \times 10^{20}$\,Mx
day$^{-1}$. The latter authors also reported on a clear relation between the cancellation
events at the PIL and an outward-directed coronal activity, but they caution
that the contribution of the cancellation events was four times smaller than
the global flux decay rate. According to this study not all flux losses could
have been attributed to cancellations at the PIL, with the resolution and
sensitivity of \textit{SOHO}/MDI magnetograms.

      \begin{figure}
       \centering
       \resizebox{\hsize}{!}{\includegraphics{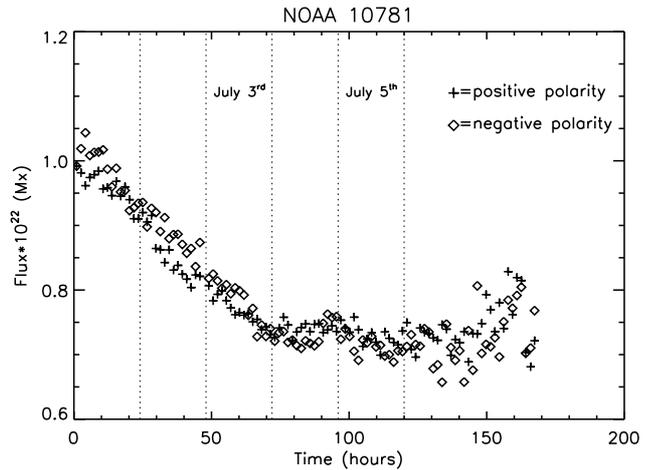}}
       \caption{Flux history curves for the positive ($+$) and negative ($\diamond$)
polarities of AR 10781. 24 hour days are separated by the vertical dotted lines.}
       \label{Fig:Fluxlosses}%
       \end{figure}

Figure \ref{Fig:Fluxlosses} displays the flux history curves for the positive
and negative polarities of this active region. The two days when TIP-II
observations were taken are highlighted in the figure. The active region boundaries
were visually selected for each \textit{SOHO}/MDI map and adapted to the
size changes of the AR originated by the well-known flux transport processes.
The magnetic flux of each pixel is multiplied by a factor
$1/\cos^2(\theta)$, $\theta$ being the heliocentric angle, to account for the
line-of-sight projection of the magnetic field (which is assumed to be
vertical) and the projection effect on the pixel area.  The positive flux of
the active region is computed by adding the contribution of all the pixels 
whose flux is larger than $-30$\,G (before multiplying by the above factor).
 Similarly, the negative
flux of the active region is computed by adding all the pixels with fluxes
below $+30$\,G. By including pixels with fluxes in the range [-30,+30]\,G (which represents
the peak-to-peak noise) in the computations, we ensure
efficient noise cancellation. The two polarities nicely coincide in magnitude
and show a similar evolution. A linear decay phase is observed during the first
three days. The flux decay rate is $9.3 \times 10^{20}$\,Mx day$^{-1}$ which is
intermediate in the range of values mentioned above. We do not persue in this work whether
this could have potentially contributed to the formation of the flux rope. Magnetic field
extrapolations of some form would be necessary to conduct such a study, and this is beyond the scope of
this paper. After 30\,\% of the observed initial flux is lost, the flux
curves show a plateau region where no more net losses or gains are seen. The 30\,\%
fractional decrease is, of course, a lower limit since we do not observe the onset
of the decay phase. This plateau region was also encountered by \citet{sainz08};
however, in their study, the amount of flux lost was found to be in the range of $50-70$\,\% of
the initial value. 

  \begin{figure*}[t!]
       \centering
       \includegraphics[width=0.9\textwidth]{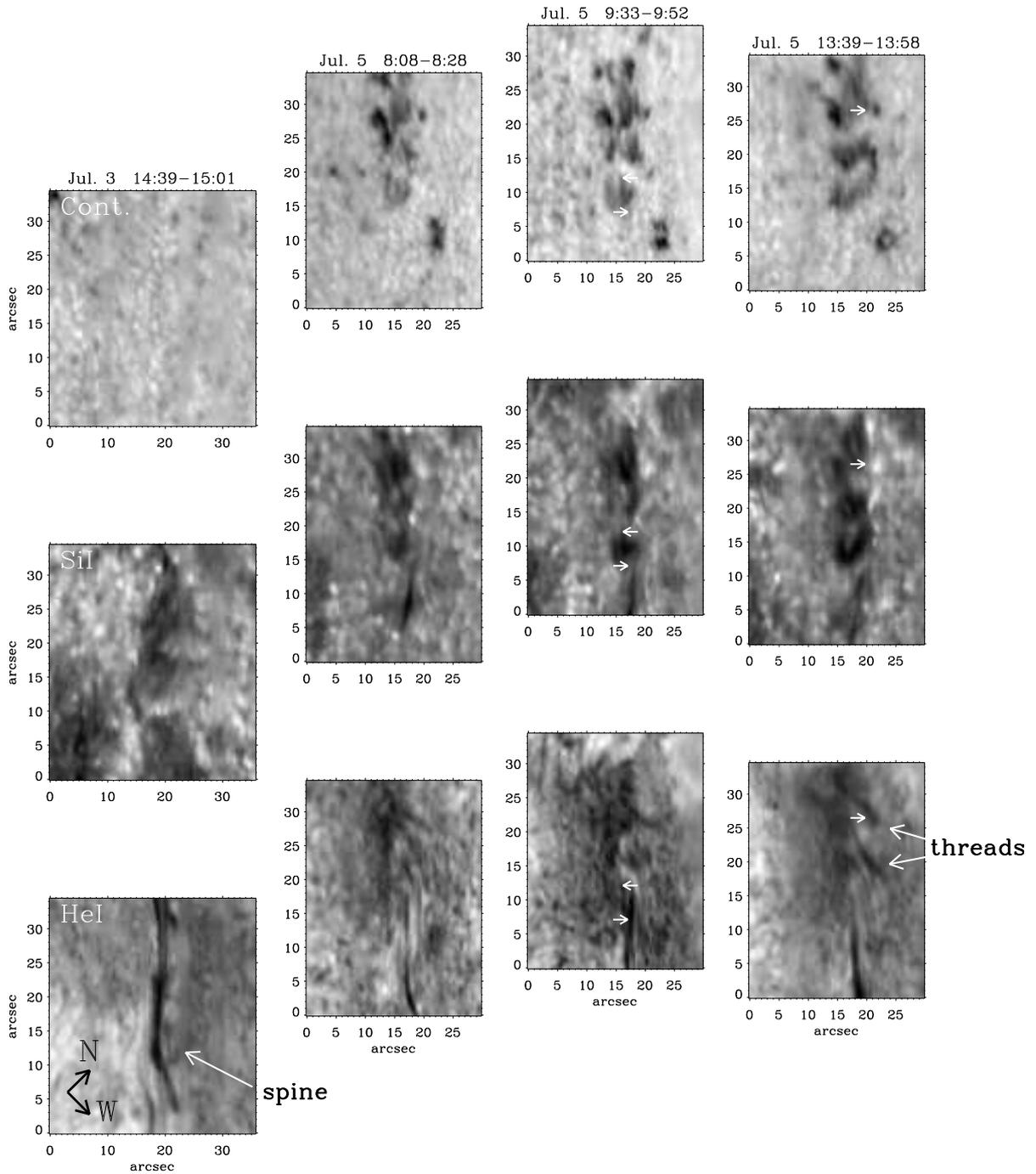}
       \caption{From \textit{left} to \textit{right} the four columns show
       slit-reconstructed images from the Tenerife Infrared Polarimeter (TIP-II) at
       different times. From \textit{top} to \textit{bottom} several
       wavelengths are presented which represent different layers in the
       atmosphere, from the photosphere to the chromosphere: continuum, \ion{Si}{i}
       10827\,\AA\ line center and \ion{He}{i} 10830\,\AA\ red core
       intensities. The panels are located at different positions to approximately
       represent the alignment of their respective FOVs. On July 3rd the filament is 
       relatively thin and no pores
       appear below it. In contrast, on July 5th big pores, orphan penumbrae 
       and \ion{He}{i} dark threads have formed in the upper half of
       the map, and the shape of the filament is rather diffuse
       as compared to the spine. Still, the spine in
       the lower part of the map is clearly seen at the same place as
       detected on July 3rd. The small white arrows indicate the position of the
       Stokes profiles presented in Figs. \ref{Fig:profhethread1},
       \ref{Fig:profspine2} and \ref{Fig:profNL3}.}
       \label{Fig:TIPmaps}%
       \end{figure*}

     \begin{figure*}[t!]
       \centering
       \includegraphics[width=0.99\textwidth]{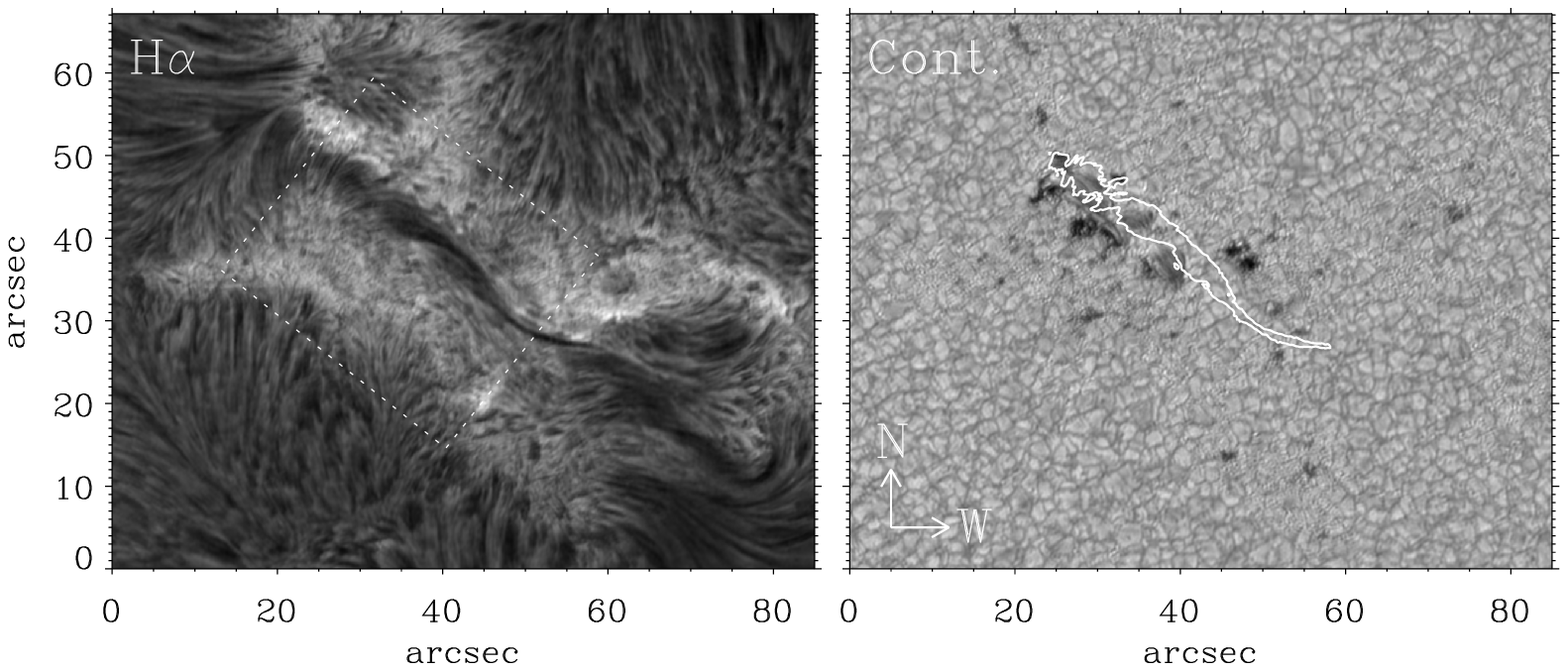}
       \caption{H$\alpha$ line core (\textit{left}) and continuum
       (\textit{right}) images from the Dutch Open Telescope taken on
       2005 July 5th at 8:44 UT. The contour of the filament is superimposed
       upon the continuum image and shows that it partially lies on top of the
       orphan penumbrae, between the pores. In the \textit{left} panel, 
       the dashed white box indicates the approximate FOV of TIP-II for
       that day.}
       \label{Fig:DOT}%
       \end{figure*}

The sliding door phenomenon was seen to occur on July 3rd. This date 
sits in the linear decay phase of the AR. However, this phase
was already taking place before and no particular slope
change is seen on July 3rd. The appearance of the orphan penumbrae
structures happens on July 4th and 5th, coinciding with the flat 
section of the flux curve. Clearly, these processes (sliding door and 
orphan-penumbrae generation) cause a minimal impact on the flux 
curves. One way to explain this result would require that the amount of 
{\em longitudinal} flux change involved in both processes were
at, or below, the noise level of the data ($\pm 0.3 \times 10^{21}$\,Mx).

\subsection{TIP-II observations}
The first set of spatial scans were taken on 2005 July 3rd, centered on the
filament. Two maps with a FOV of $36\arcsec \times 35\arcsec$, which was
unfortunately not large enough to cover the whole active region (see Fig.
\ref{Fig:MDIevol}), were acquired with TIP-II between 13:53 and 15:01 UT. The
filament is seen all along the vertical direction.
Slit-reconstructed maps at different wavelengths for one of the two data sets 
for this day are
presented in the first column of Fig. \ref{Fig:TIPmaps}. The   
frames in this figure are located at different positions to approximately
represent the alignment of their respective FOVs. The \textit{bottom
left} panel shows a tight filament spine, inferred from the strong \ion{He}{i}
absorption, which extends along the vertical direction. There are no 
big pores, but only some small magnetic features (dark patches) 
seen among the granulation in the continuum image on the \textit{top left} panel
of the figure. The \ion{Si}{i} line core image (in the
middle row) shows dark areas with larger absorption in regions of weak  
longitudinal field (this line weakens in faculae, as do most
photospheric lines).

The second set of spectropolarimetric data was taken on July 5th between 7:36
and 14:51 UT, columns 2--4 in Fig. \ref{Fig:TIPmaps}. Each scan took around 20
minutes to complete. The acquired maps were centered at a highly interesting
area with very tight opposite polarities that corresponded,
in continuum image, to pores and orphan penumbrae, all located
along the PIL. Note that these maps overlap with the upper half of the former
map from July 3rd (see black boxes in Fig. \ref{Fig:MDIevol} to get a better
notion of the FOV). The \ion{He}{i} red core intensity images in the
\textit{bottom} row of Fig. \ref{Fig:TIPmaps} still show the spine of the
filament in the lower part. Moreover, the filament in the upper part appears to
be more diffused and extended. One can easily distinguish dark Helium
threads formed on July 5th especially in the \textit{lower right} panel of
Fig. \ref{Fig:TIPmaps}. Many authors have observed the presence of threads in
filaments and prominences before
\citep[e.g.,][]{menzel60,engvold76,lin05,lin08,okamoto07}.  It is generally
believed that dark thin features in the chromosphere near ARs trace magnetic
field lines. This idea is particularly interesting when applied to threads seen in AR
filaments, since it could explain the presence of magnetic dips where plasma is
trapped. Nevertheless, care must be taken, since a recent paper by
\citet{jaime11} proved that chromospheric fibrils mostly, but not always, trace
magnetic field lines. From our maps we see that the threads observed with
TIP-II change only slightly in a time range of 5--6 hours. This becomes apparent
when closely comparing the Helium maps of columns 2 and 4 in Fig.
\ref{Fig:TIPmaps}. The short white arrow in the \textit{bottom} panel of the last column points at 
one of these threads, and can be compared with the same position in the
\ion{Si}{i} core and continuum images. From these maps we see that the most
prominent \ion{He}{i} threads are located above pores or orphan penumbrae. 
A more detailed study of the magnetic configuration of these threads is presented
below in Sect. \ref{Sect:threads}.

Strong absorption in the \ion{Si}{i} line core is present below the spine of the filament on July
5th (see Fig. \ref{Fig:TIPmaps}). 
It is remarkable that the axis of the filament seems to lay so low in 
the atmosphere that even the highest layers of the photospheric Silicon absorption 
line (at the core of the line) trace it. 

\subsection{DOT observations}
High resolution (0\farcs071/px) H$\alpha$ 6563\,\AA\ images from the 
Dutch Open Telescope \citep[DOT;][]{DOT}  for the morning of 2005 July 5th
confirm the presence of the filament, which has an inverse
S-like shape and a spine in its lower part. This is also
confirmed by inspection of \textit{TRACE} at $171$\,\AA\ images.
Such an inverse S-like shape is likely to be expected in the northern 
hemisphere \citep{pevtsov01}. The \textit{left} panel of
Fig. \ref{Fig:DOT} shows one image of the H$\alpha$ data set taken by the DOT
at 8:44 UT. The filament is surrounded by a bright plage. The image reveals
small arch-like structures in the upper half of the filament that are almost
perpendicular to its axis and, in the spine, then stretch along it towards its center. These
archs can be identified as the H$\alpha$ counterpart of the \ion{He}{i}
threads mentioned before. A continuum image (with a 3\,\AA\ bandpass) is
shown in the \textit{right} panel of Fig. \ref{Fig:DOT}. This image is spatially
aligned with the H$\alpha$ panel on the \textit{left} and has, superimposed on it in white,
the H$\alpha$-contour of the filament. It is clear that the broader part of the
filament lies on top of the orphan penumbrae, in between the pores, whereas
the spine in the lower part is only surrounded by weaker magnetic features
(small gray patches in the continuum image). 

\section{Spectral line inversion and data analysis} \label{Sect:inversions}

The average signal-to-noise ratio after binning the data is about 2000.
Since the polarization signals of the \ion{Si}{i} 10827\,\AA\ line are
much stronger than those of the \ion{He}{i}
triplet, it is not necessary to apply the same binning to the
\ion{Si}{i} Stokes profiles and suffer the corresponding
loss of spatial resolution. However, for a better comparison between them we
use, unless stated otherwise, the same binning criterion for both.
Two different inversion
codes were used to fit the observed Stokes profiles. 
Both use a nonlinear least-squares Levenberg--Marquard
algorithm to minimize the differences between the observed and synthetic 
spectra. For the \ion{He}{i} 
10830\,\AA\ triplet a Milne--Eddington (ME) inversion code
\citep[MELANIE;][]{socas01} which computes the Zeeman-induced Stokes spectra in
the incomplete Paschen-Back (IPB) effect regime was used \citep[see][for a detailed
study of the effects of the IPB in the \ion{He}{i} multiplet]{socas04}. MELANIE
does not take into account the atomic-level polarization due to anisotropic
radiation pumping and its modification via the Hanle effect. Nevertheless, 
by comparing the output of three different techniques with different levels of
physical complexity, \citet{kuckein09} showed that the polarization signals in this 
AR filament are dominated by the Zeeman
effect, fact which was supported by the excellent performance of the MELANIE inversions. 
This code normally uses a set of eleven free parameters which 
are modified in order to obtain the best fits to the observed Stokes profiles: 
magnetic field strength ($B$), inclination ($\gamma$), azimuth ($\phi$), line strength
($\eta_0$), Doppler width ($\Delta\lambda_\mathrm{D}$), a damping parameter,
LOS velocity ($v_\mathrm{LOS}$), the source function at $\tau = 0$ and its
gradient, macroturbulence factor and stray-light fraction ($\alpha = 1 - f$),
where $f$ is the filling factor -- or fraction of the magnetic component which
occupies the resolution element. MELANIE requires an initial guess
for the free parameters. For this purpose we carried out a
magnetograph analysis and a gaussian fit to Stokes $I$ to obtain 
rough estimates for $B$, $\gamma$, $\phi$,
$\Delta\lambda_\mathrm{D}$ and $v_\mathrm{LOS}$ for each point along the slit.

For the thermodynamical parameters, several inversions of a 
few representative cases with random initial
values, but within reasonable boundaries, were carried out. 
Averages of the retrieved parameters were calculated and
used as initial guesses for the rest of the inversions. 

Test inversions were carried out, where
$\alpha$ was left as a free parameter, to control the fraction of a given 
non-magnetic stray-light profile in the inversions. 
The results yielded almost no changes in the retrieved inclinations,
azimuths and LOS velocities. The inferred magnetic field strengths
were larger, but never by more than 100\,G, than in the case
without stray-light. Therefore, we finally
decided to fix this parameter for the \ion{He}{i} ME inversions, 
setting the filling factor to 1 (i.e., $\alpha = 0$). The macroturbulence
factor was also fixed to the calculated theoretical value of the spectral resolution
of the data -- a combination of slit, grating and pixel resolution, which yielded $1.2$\,
km\,s$^{-1}$. Thus, only 9 free parameters were left in the MELANIE fit.

       \begin{figure*}[t!]
       \centering
       \includegraphics[width=0.8\textwidth]{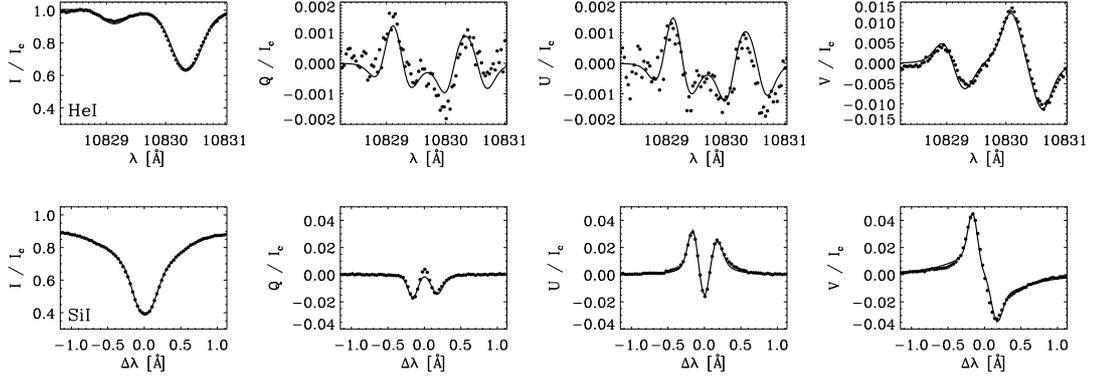}
       \caption{Stokes profiles of the \ion{He}{i} 10830\,\AA\ triplet (four
       \textit{top} panels) and the \ion{Si}{i} 10827\,\AA\ line (four
       \textit{bottom} panels) at coordinates $x \sim 20$\arcsec\ and $y \sim
       $26\farcs5 in Fig. \ref{Fig:TIPmaps}, on July 5th, 13:39--13:58 UT
       column (see white arrows). The selected point is located inside the dark
       \ion{He}{i} threads. The dots represent the observed profiles which were
       binned in order to obtain a larger S/N ratio. The solid line
       corresponds to the best fit achieved with a Milne-Eddington / LTE
       inversion code for the \textit{upper / lower} panels. Stokes $Q$ and $U$
       present the typical Zeeman-effect three-lobe signature. For Helium the
       abscissae indicates the wavelength while for Silicon it shows the
       distance to the \ion{Si}{i} line core center in Angstrom units. The fit
       of the Helium Stokes profiles provided the following magnetic parameters
       $B_{\mathrm{He}} = 592 \pm 31$\,G, $\gamma_{\mathrm{He}} = 69.6^\circ
       \pm 0.8^\circ$ and $\phi_{\mathrm{He}} = 115.5^\circ \pm 0.3^\circ$, while
       the Silicon inversion gave $B_{\mathrm{Si}} = 1185 \pm 22$\,G,
       $\gamma_{\mathrm{Si}} = 71.7^\circ \pm 1.1^\circ$ and
       $\phi_{\mathrm{Si}} = 53.7^\circ \pm 1.3^\circ$. A filling factor of $f
       \sim 0.7$ was inferred from the SIR inversion.}
       \label{Fig:profhethread1}%
       \end{figure*}

       \begin{figure*}[t!]
       \centering
       \includegraphics[width=0.8\textwidth]{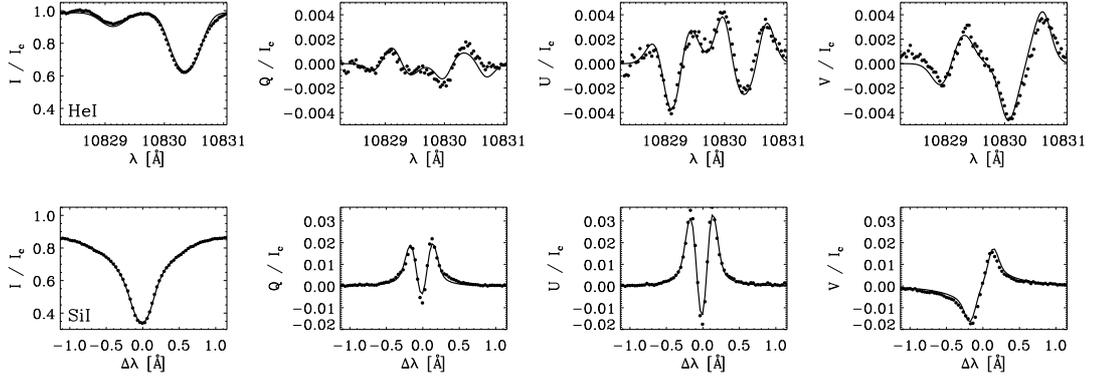}
       \caption{Same as Fig. \ref{Fig:profhethread1} but for coordinates $x
       \sim 17$\arcsec\ and $y \sim 7$\arcsec\ in Fig. \ref{Fig:TIPmaps}, on
       July 5th, 9:33--9:52 UT column (see white arrows pointing from
       \textit{left} to \textit{right}). These profiles belong to the spine of
       the filament. The magnetic field obtained from this particular fit for
       Helium was $B_{\mathrm{He}} = 788 \pm 11$\,G, $\gamma_{\mathrm{He}} =
       95.0^\circ \pm 0.2^\circ$ and $\phi_{\mathrm{He}} = 54.3^\circ \pm
       1.1^\circ$ and for Silicon $B_{\mathrm{Si}} = 1058 \pm 26$\,G,
       $\gamma_{\mathrm{Si}} = 99.5^\circ \pm 0.8^\circ$ and
       $\phi_{\mathrm{Si}} = 32.2^\circ \pm 1.5^\circ$ with a $f \sim 0.8$. }
       \label{Fig:profspine2}%
       \end{figure*}

      \begin{figure*}[t!]
       \centering
       \includegraphics[width=0.8\textwidth]{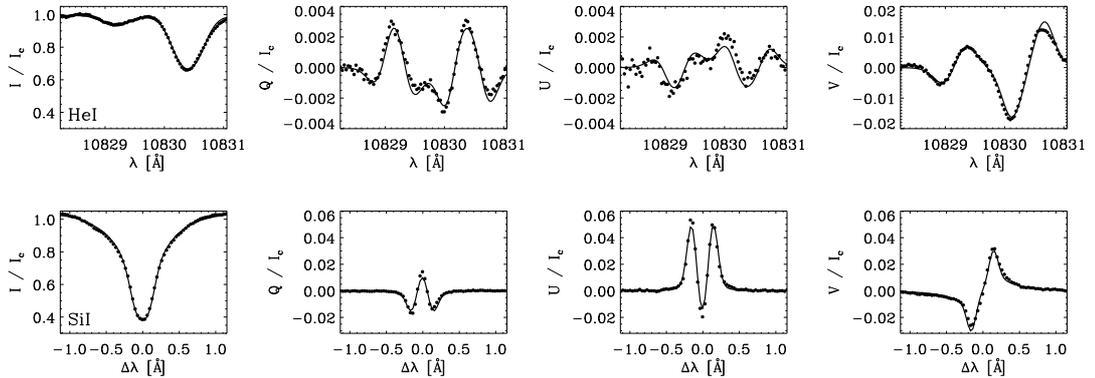}
       \caption{Same as Fig. \ref{Fig:profhethread1} but for coordinates $x
       \sim 16$\arcsec\ and $y \sim 12$\arcsec\ , see white arrows in Figures
       \ref{Fig:TIPmaps} (arrow pointing from \textit{right} to \textit{left}), 
       \ref{Fig:magnetoHe} and \ref{Fig:magnetoSi}, on July
       5th, 9:33--9:52 UT map. The Stokes profiles were observed at the
       polarity inversion line. The fitted parameters for the Helium are
       $B_{\mathrm{He}} = 878 \pm 11$\,G, $\gamma_{\mathrm{He}} = 109.8^\circ
       \pm 0.3^\circ$ and $\phi_{\mathrm{He}} = 76.2^\circ \pm 1.8^\circ$ and
       for the SIR case are $B_{\mathrm{Si}} = 806 \pm 21$\,G,
       $\gamma_{\mathrm{Si}} = 99.9\circ \pm 0.8^\circ$, $\phi_{\mathrm{Si}}
       = 56.1^\circ \pm 1.2^\circ$ and a photospheric filling factor of $f \sim 0.8$. }
       \label{Fig:profNL3}%
       \end{figure*}

The photospheric \ion{Si}{i} 10827\,\AA\ line was inverted using the SIR inversion
code \citep{SIR} which is based on spectral line response functions (RFs) and assumes
Local Thermodynamical Equilibrium (LTE) and hydrostatic equilibrium to 
solve the radiative transfer equation. 
The advantage of using the SIR code is that a
depth-dependent stratification of the physical parameters as 
a function of the logarithm of the LOS continuum optical
depth at $5000$\,\AA\ ($\log\tau$) can be obtained. 

We tested the effects on the inversions of using different initial 
model atmospheres (umbra, penumbra and plage models), to asses
their performance when reproducing the observed Stokes profiles. 
The penumbral model from
\citet{toro94} seemed to yield the best results, which is consistent with what
we see in the observations at the photosphere. However, some modifications 
to the model (such as assuming an initial constant 
magnetic field strength of 500\,G  and a LOS velocity of 0.1
km\,s$^{-1}$) had to be implemented. 

Different initial values for the inclination and azimuth did not affect 
the final fit to the observed Stokes profiles. 
Macroturbulence was fixed to the same value as for the ME inversions. 
However, for the inversion of the \ion{Si}{i} line, the stray-light was 
initialized with $30\%$ and left as a free parameter in the fit. 
Stray-light profiles for each map were computed by averaging the Stokes $I$
of non-magnetic areas, i.e., regions where Stokes $Q$, $U$ and $V$ are negligible.
The necessary atomic data for the \ion{Si}{i} 10827\,\AA\ line was 
taken from the work of \citet{borrero03}. In particular, the value of the 
logarithm of the oscillator strength times the multiplicity of the lower level 
that we used was $\log (g f) = 0.363$.

To have a rough idea of the formation height of Silicon in order to associate an 
appropriate optical depth for the inferred vector magnetic field, response functions to
magnetic field perturbations 
at various positions near or in the filament were calculated.
Our atmospheric model covers heights from 1.2 up 
to -4.0 (in $\log\tau$ units). The largest sensitivity for both days was found to
take place at $\log\tau = -2$. Thus, from this point onwards, all of the 
figures derived from the inversions of the \ion{Si}{i} line are 
referred to this height.

Figures \ref{Fig:profhethread1} -- \ref{Fig:profNL3} show the results of the MELANIE
and SIR inversions of the Stokes parameters for three different positions
along the filament: one in a Helium dark thread, one in the spine and one at the PIL. 
Each figure is made of 8 plots: in the \textit{top (bottom)} row we present the \ion{He}{i}
10830\,\AA\ (\ion{Si}{i} 10827\,\AA) observed Stokes $I$, $Q$, $U$ and $V$ 
profiles (dots) obtained after performing the binning and the best fit achieved
by the inversion code (solid line). The exact location of the selected fits is
indicated by short white arrows in Figs. \ref{Fig:TIPmaps}, \ref{Fig:magnetoHe} and
\ref{Fig:magnetoSi} (see the captions of the figures for a detailed explanation). 
MELANIE does not provide the uncertainties in the retrieved atmospheric parameters. 
To estimate them, we used the synthetic Stokes profiles which resulted from 
the best fit of the model to the data. Then, several different realizations of the
noise (with an amplitude of that of the noise in the observations), were added to
the synthetic Stokes profiles, which were in turn inverted again. 
The standard deviation computed from the spread in the values of the retrieved 
parameters provided the errors quoted in the captions of Figs.
\ref{Fig:profhethread1}, \ref{Fig:profspine2} and \ref{Fig:profNL3}. The SIR
inversions directly provide uncertainties which are proportional to the inverse of
the response functions to changes in the physical parameters.

We found a very satisfactory performance of the ME inversion code when 
fitting the \ion{He}{i} Stokes profiles. This has already been described by
\citet{kuckein09} who found in these data a ubiquitous presence of Zeeman-like
signatures (see Stokes $Q$ and $U$ frames in Figs. \ref{Fig:profhethread1} --
\ref{Fig:profNL3}) with no apparent contribution of atomic level polarization
and its modification by the Hanle effect. The presence of strong Stokes $Q$ and
$U$ profiles is indicative of strong transverse magnetic fields in the filament.  
In the photospheric \ion{Si}{i} 10827\,\AA\ absorption line we also find strong Stokes
$Q$ and $U$ profiles, which occasionally have larger amplitudes than 
the corresponding Stokes $V$. Consequently, a predominant horizontal field is 
found at photospheric heights too.

\subsection{NLTE \ion{Si}{i} line formation}
Prior to the inversions of the \ion{Si}{i} 10827\,\AA\ line,
we studied the possible impact of 
non--local thermodynamic equilibrium (NLTE) effects in 
the retrieved atmospheric parameters. SIR synthesizes the Stokes profiles under the assumption of 
local thermodynamic equilibrium. A recent study by \citet{bard08} reported that
this line can be significantly affected by NLTE conditions. The authors showed that the
NLTE \ion{Si}{i} line core intensity is deeper than the analogous LTE result for two
different model atmospheres: quiet sun (FALC model) and sunspot umbra models
(SPOTM model). To study possible effects on our inversion results, we introduced the
departure coefficients $\beta$ (which are defined as the ratio between population
densities in NLTE over LTE) from the above cited paper (kindly provided
to us by M. Carlsson), into the SIR code. Inversions of a few
selected cases where run with the $\beta$ coefficients from these models (together with the
LTE case) and the differences between the inferred magnetic
field strength ($B$), inclination ($\gamma$) and azimuth ($\phi$) 
were studied. No significant changes were
found for $\gamma$ and $\phi$ between the $\beta$ and non-$\beta$ inversions,
however, the field strength presented some variations. Differences 
with an rms value of $100$\,G, in the spine, and of $150$\,G, in the diffuse filament 
region in the upper part of the maps, were found. We also studied the behavior of the LOS
velocities taking into account the various options for the $\beta$ coefficients. 
The rms changes found in the spine and in the orphan penumbrae were of
around $0.1$ and $0.2$\,km\,s$^{-1}$, respectively.
 
These values are small, similar to other errors arising from photon
noise or systematics from the velocity calibration. NLTE effects of the \ion{Si}{i} 10827\,\AA\
line are certainly non-negligible when estimating the temperature ($T$) stratification,
which was found in our tests, but this has no impact on our purely magnetic study. 
In general, the temperatures above the range  of $\log\tau = -0.5$ and $-1.0$ do depend 
drastically on the departure coefficients used and cannot be trusted unless a self consistent
NLTE inversion of the line is carried out. However, we emphasize that no major influence on
the returned vector magnetic field and velocities was found when neglecting NLTE departue 
coefficients. 

\begin{figure*}[t!]
       \centering
       \includegraphics[width=0.95\textwidth]{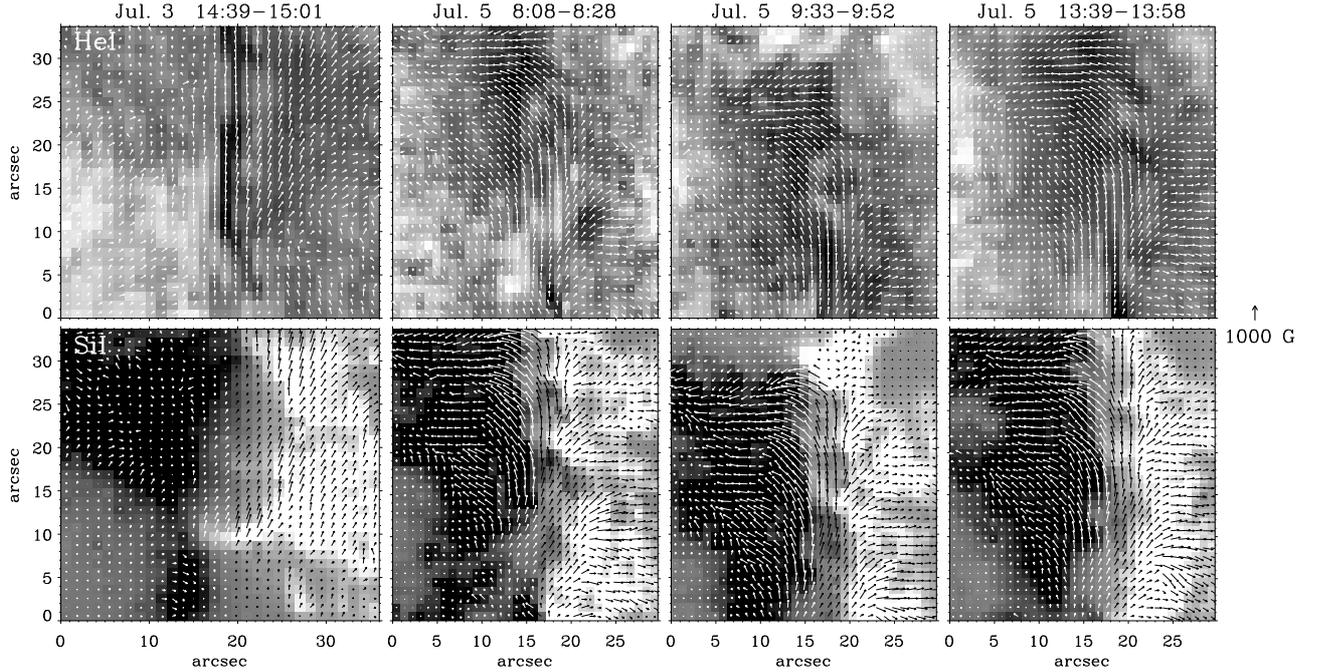}
       \caption{\textit{Top}: Four gray-scale slit reconstructed images
       representing \ion{He}{i} red core intensity, increasing with time from
       left to right. Superimposed are white arrows indicating
       the horizontal magnetic fields observed with TIP-II in the \ion{He}{i}
       10830\,\AA\ multiplet after solving the 180$^\circ$ ambiguity.
       Especially in the \textit{top right} panel the magnetic field lines are
       well aligned with the dark \ion{He}{i} threads. \textit{Bottom}:
       Background images show photospheric vertical fields
       $B_\mathrm{vert}^\mathrm{Si}$ observed with TIP-II in the \ion{Si}{i}
       10827\,\AA\ spectral line. The images are saturated at $\pm400$\,G to
       emphasize the PIL. The black and white arrows (depending on the
       background for a better contrast) represent the horizontal fields. Note
       that the arrows are parallel to the polarity inversion line which lies
       in the gray area between 15 and 20 arcseconds on the $x$-axis and
       extends from the \textit{lower} to the \textit{upper} part of each panel. }
       \label{Fig:vector5jul}
       \end{figure*}

\subsection{180$^\circ$ ambiguity}
Before transforming the vector magnetic field from the line-of-sight into the
local solar frame of reference, we need to solve the
well-known 180$^\circ$ ambiguity. Due
to the angular dependence of Stokes $Q$ and $U$ with the azimuth, two configurations
for the magnetic field, compatible with the measured Stokes profiles, 
are obtained. Hence, the correct solution must be found by
other means. Several methods for resolving the ambiguity in the
azimuth can be found in the literature \citep[][and references
therein]{martin08}. However, we were not able to apply them to our filament
since, neither did the H$\alpha$ high resolution images clearly show barbs, nor
was a sunspot located immediately next to the filament to assure the continuity of the
azimuth. Instead, we used the interactive IDL utility AZAM \citep{lites95} to minimize the
discontinuities between the two solutions of the azimuth 
on a pixel-by-pixel basis. Once
we obtained the discontinuity-removed maps of the azimuth and the inclination
in the local solar frame, we were left with two global solutions for each map,
one of which needed to be rejected. 
In one of the two solutions the vector magnetic field pointed  
in the N-E direction (upward in Fig. \ref{Fig:TIPmaps}), whilst in the other
solution it pointed S-W (downward in the same figure). In order to
select one or the other, we invoked the large scale magnetic topology
of the active region. Since our AR appeared during Solar Cycle 23 and was located
in the northern hemisphere, the leader polarity must correspond to a 
magnetic field pointing outwards from the solar surface while the 
following polarity must contain a field that points inwards. In such a
configuration, the toroidal field lines that originated the AR 
necessarily had to point from West to East. If we now acknowledge that
the leading polarity leans slightly towards the equator and
the following polarity towards the pole  (Joy's Law), this generates
a poloidal component pointing North, which corresponds to
a N-E orientation of the vector magnetic field on Fig. \ref{Fig:TIPmaps} 
(pointing upwards) as the most probable solution.

\section{Results}
\subsection{Vector magnetic field analysis in and underneath the filament}

Average horizontal and vertical magnetic fields in each pixel were
calculated using the equations given by \citet{landi92}:

\begin{equation} \label{eq:Bparallel}
 B_\mathrm{vert} = f ~\mathbf{|B|} ~\cos\gamma,
\end{equation}

\begin{equation}\label{eq:Bperp}
 B_\mathrm{hor} = \sqrt{f} ~\mathbf{|B|} ~ \sin\gamma,
\end{equation}
where $f$ is the inferred filling factor, $\mathbf{|B|}$ is the total field
strength and $\gamma$ is the inclination with respect to the local vertical.
As mentioned in Sect. \ref{Sect:inversions} we assumed $f = 1$ for Equations
\ref{eq:Bparallel} and \ref{eq:Bperp} in the case of the Helium inversions.
Since $\gamma$ is the inclination in the local reference frame, vertical magnetic 
fields are radially oriented whereas horizontal magnetic fields are parallel to the 
surface.

Figure \ref{Fig:vector5jul} shows arrows denoting the horizontal magnetic
field inferred from the \ion{He}{i} triplet (\textit{upper} panel) and 
the \ion{Si}{i} line (\textit{lower} panel) inversions after solving the
180$^\circ$ ambiguity for both days. Arrows are in black or white for a
better distinction from the background. The gray-scale image in the
\textit{top} and \textit{bottom} panels correspond, respectively, to a 
slit-reconstructed map of the \ion{He}{i} core intensity (for the red component)
and a photospheric magnetogram (saturated at $\pm 400$\,G).  
For July 3rd, the thin filament axis (or spine) lies
mainly on top of the broad polarity inversion line, i.e., the gray area between
opposite polarities.  The photospheric vector magnetic field arrows in the
\textit{lower} panel depict an organized field with a
predominantly inverse configuration
at the PIL, i.e., the component of the transverse
field perpendicular to the PIL has a preference to point 
from negative (black) to positive (white)
polarity. Field vectors have, typically, a $\sim 45^\circ$
orientation clockwise with respect to the filament axis seen
in the \textit{upper} panel. Note also that the transverse magnetic fields 
are stronger (longer arrows) to the right (positive polarity) than to the left 
(negative polarity) of the PIL.  
The chromospheric horizontal magnetic field presented in the \textit{upper}
panel appears to be highly sheared (i.e., field lines are parallel to the PIL
and to the filament axis), but also displays an inverse configuration which
dominates the surroundings of the spine (main axis). It is important to 
emphasize here that the chromospheric fields in the filament axis or spine
have a stronger shear than the photospheric ones, the latter showing a
preference for an inverse configuration.

The last three columns of Fig. \ref{Fig:vector5jul} present the vector 
magnetic field topology of the
maps of July 5th using the same criterion as described above. The spine is also
seen on this day at the bottom of the panels, but the top part is dominated
by a more diffuse filament in the chromosphere and by pores and orphan penumbrae
in the photosphere. In all
Helium images (\textit{upper} panels), the longest arrows, i.e, larger
horizontal field component, are found in the filament between 10\arcsec\ and
22\arcsec\ on the $x$-axis. This also happens at photospheric layers. This
readily shows that the filament area harbors the strongest horizontal fields
in the region at both heights.
Changes of the vector magnetic fields through the day are almost
insignificant in a 5--6 hour time scale. 
Interestingly the field lines are aligned with the dark Helium threads (top panels). 
This can be easily seen in the last column of Fig. \ref{Fig:vector5jul}. 

The inferred chromospheric field is clearly aligned with the spine (sheared
configuration) in the lower part of the images and has a slightly inverse
orientation close to it. This is the same behavior as observed on July
3rd. The photospheric fields show a predominance of the inverse
configuration all over the spine region which, again, coincides with what had
been observed two days earlier. In the \textit{upper} part of the \ion{He}{i} intensity
images, above the spine, the chromospheric horizontal fields smoothly change orientation,
displaying a normal configuration. 
This shift in orientation is complete on the top part of the
figures, where the fields perpendicular to the PIL point
straight from positive to negative polarity. The
\textit{bottom} panels of Fig. \ref{Fig:vector5jul}\, for this day, show
that the photospheric horizontal fields are more aligned with the
PIL, showing a more sheared configuration than their chromospheric
counterparts.

Thus, the filament seems to be divided in two substructures: (1) a 
chromospheric spine with field lines parallel to its axis and an inverse 
polarity configuration in the photosphere below it and (2) extensive
dark Helium patches or threads which show a normal polarity configuration 
above and, sheared field lines running along the PIL underneath, in the 
photosphere.

\begin{figure*}[t!]
       \centering
       \includegraphics[width=0.99\textwidth]{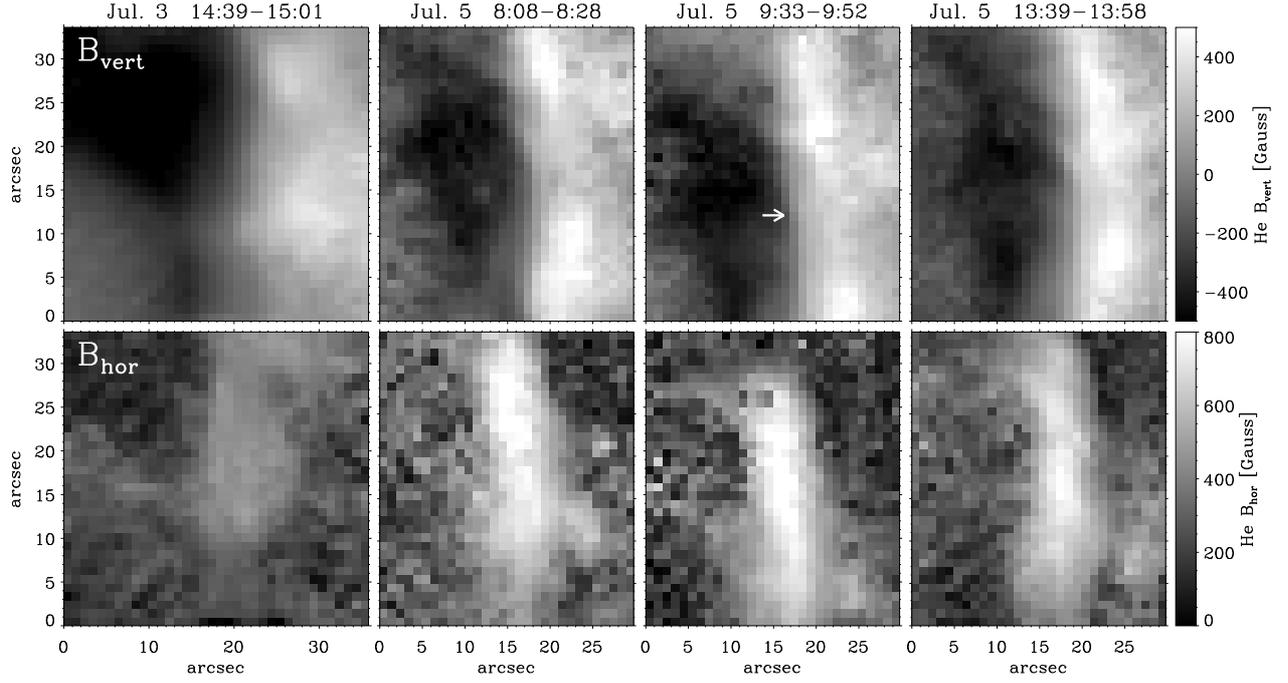}
       \caption{The gray-scale images indicate the two components of the magnetic field strength
       inferred from the ME inversions of the \ion{He}{i} 10830\,\AA\ triplet.
       The \textit{upper} / \textit{lower} four panels show the vertical
       ($B_\mathrm{vert}$) / horizontal ($B_\mathrm{hor}$) fields in the
       local frame of reference. All images in the same row have the same intensity scale.
       Vertical and horizontal fields are saturated at $\pm 500$\,G and $800$\,G respectively.
       Both polarities are close together and the PIL is only a few arcseconds wide in all panels.
       The white arrow indicates the location of the Stokes profiles shown in Fig. \ref{Fig:profNL3}. }
       \label{Fig:magnetoHe}%
       \end{figure*}

\subsection{Helium threads}\label{Sect:threads}
We performed a more detailed study of the chromospheric \ion{He}{i} threads
which appear in the \textit{top right} panel of Fig. \ref{Fig:vector5jul}. The
centers of two long threads are located around $[x,y] = [20\arcsec, 28\arcsec]$
and $[20\arcsec, 20\arcsec]$, herafter thread1 (TH1) and thread2 (TH2),
respectively. They extend from just above the PIL to far into the positive polarity region.
According to the orientation of the field lines, which are
aligned with the threads, a normal polarity configuration pointing from
positive to negative is present and therefore it is
natural to consider them as simple arches anchored in one polarity and
reaching over the other. 
We suggest that these threads act like footpoints of the filament. The following inferred
magnetic properties are consistent with this interpretation. Along thread1,
between $x \in [18\arcsec,22\arcsec]$, we found that the inclination
changed from a more horizontal configuration, $\gamma_\mathrm{TH1} \sim 66^\circ$, 
near the PIL to a more vertical one, $\gamma_\mathrm{TH1} \sim 35^\circ$ 
further away from it, in the positive polarity plage region (an inclination of 
$\gamma = 90^\circ$ means
that the field lines are parallel to the solar surface).
Along with this, the horizontal fields decreased from $B_\mathrm{hor}^\mathrm{He} \sim 570$
to $360$\,G whereas the total field strength remained almost constant along the structure, i.e.,
the vertical component of the field $B_\mathrm{vert}^\mathrm{He}$ increased. The same
happened in thread2, between $x \in [19\arcsec,22\arcsec]$: inclinations became
more vertical away from the PIL, going from $\gamma_\mathrm{TH2} \sim 68^\circ$ to $38^\circ$, 
while the horizontal fields decreased from $B_\mathrm{hor}^\mathrm{He} \sim 620$ to $360$\,G.
These findings suggest that these threads magnetically link the chromosphere and
the photosphere. \ion{Si}{i} data also show that the fields are more vertical
near the ending points of both threads.
As a consequence, the threads could constitute a channel
for plasma to flow down along them. Mass flows from the filament into the photosphere
through these channels will be studied in the second paper of this series.

\subsection{Vertical and horizontal magnetic field components}

\begin{figure*}[t!]
       \centering
       \includegraphics[width=0.99\textwidth]{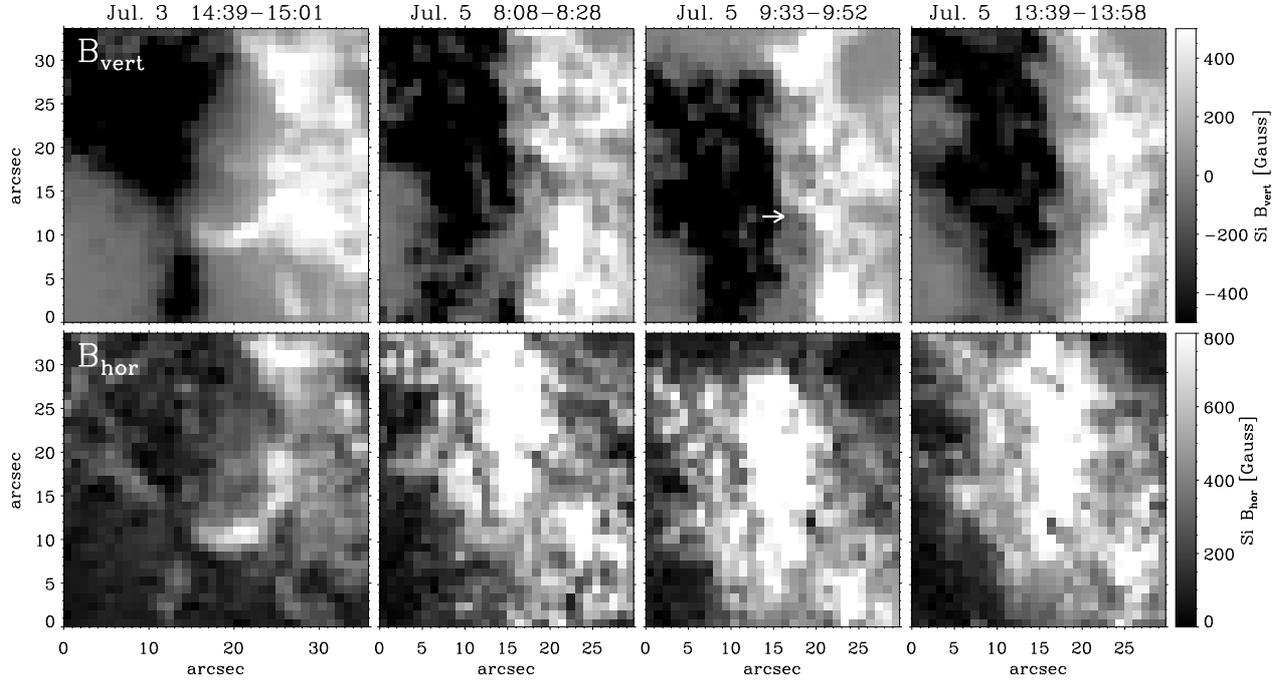}
       \caption{Same as Fig. \ref{Fig:magnetoHe} but for the SIR inversions of
       the \ion{Si}{i} 10827\,\AA\ line. On July 5th, the horizontal fields are concentrated
       at the pores (better seen when compared to Fig. \ref{Fig:TIPmaps}) and
       reach top values of up to 1100\,G. Black (negative) and white
       (positive) polarities of the \textit{top} panels are close together
       indicating a very compact AR.  } \label{Fig:magnetoSi}%
       \end{figure*}

The changes in the vertical ($B_\mathrm{vert}$) and horizontal ($B_\mathrm{hor}$) magnetic
fields at both heights and for both days 
are displayed in Figures \ref{Fig:magnetoHe} and \ref{Fig:magnetoSi}.  
Several properties are worth mentioning: 

\begin{enumerate}

\item  A comparison between Figs. \ref{Fig:magnetoHe} and \ref{Fig:magnetoSi}
shows that the horizontal fields at the spine of the filament are stronger in
the chromosphere than in the photosphere underneath. This is consistent throughout
both days, along the spine on July 3rd and outside the pores and
orphan penumbrae (the lower half of the maps) on July 5th.  On
average, $B_\mathrm{hor}^\mathrm{He} > B_\mathrm{hor}^\mathrm{Si}$ by $\sim 100$\,G. 

\item 
Fig. \ref{Fig:magnetoHe} shows that the spine in the chromosphere has a weaker
horizontal magnetic field ($B_\mathrm{hor}^\mathrm{He}\sim$ 400--500\,Gauss) than 
the region above the orphan penumbrae observed on July 5th,
which reaches horizontal field strengths as high as 800\,G
\citep[this was the region studied in detail by][]{kuckein09}.

\item The aforementioned chromospheric strong fields of July 5th are 
clearly related to the presence of pores and orphan penumbrae in the photosphere.
The corresponding photospheric horizontal fields are the strongest there too 
(see Fig. \ref{Fig:TIPmaps} for the exact location of the pores and orphan penumbrae). 
The observed $B_\mathrm{hor}^\mathrm{Si}$ in these areas is in the
range of 1000--1100\,G. 

\item 
Vertical fields seem to be very similar at both heights. However,
$B_\mathrm{vert}^\mathrm{He}$ appears to be weaker and have a more homogeneous distribution 
than $B_\mathrm{vert}^\mathrm{Si}$.

\end{enumerate}

It is of interest to see the spatially widening and narrowing between opposite
polarities. On July 3rd there is a big gap (gray area of around 6\arcsec\ wide)
between positive and negative polarities in the $B_\mathrm{vert}^\mathrm{Si}$
image. Interestingly, the absence of vertical fields is better seen at
photospheric heights (see \textit{top left} panel of Fig. \ref{Fig:magnetoSi})
rather than in the chromosphere. The same widening can also be seen in the
MDI LOS magnetograms (see
Fig.  \ref{Fig:MDIevol}; starting at the second row from the \textit{top}). Two
days later, on July 5th, the opposite polarities have moved closer together
forming an extremely compact active region PIL. Although only the upper half map of July 3rd
overlaps with the lower half of July 5th, the gap is undoubtly smaller on the second day.

\subsection{Azimuth changes between the photosphere and the chromosphere}
\begin{figure*}[t!]
       \centering
       \includegraphics[width=0.99\textwidth]{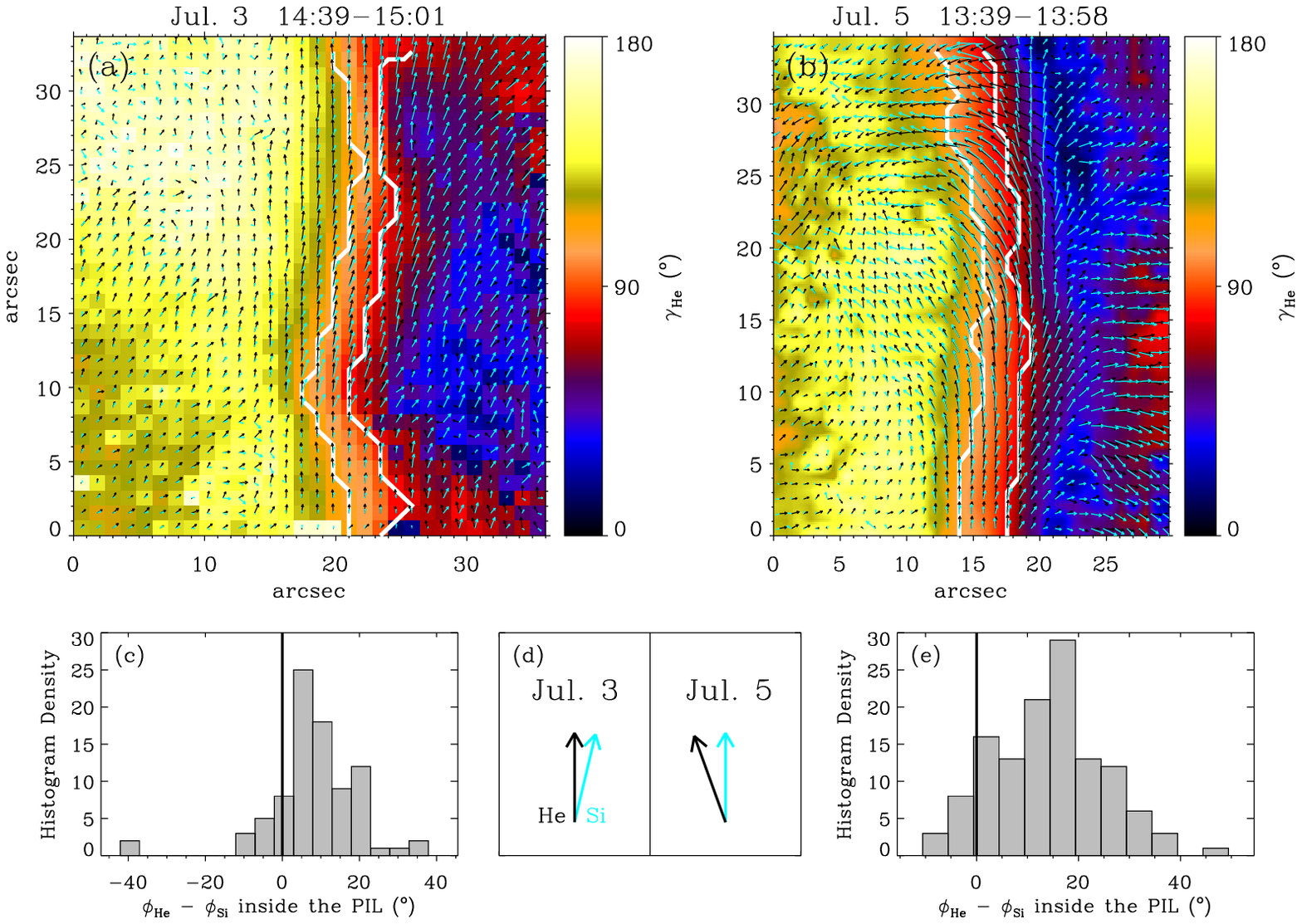}
       \caption{(a--b) Background color image shows the inclination angle
       $\gamma_\mathrm{He}$ inferred from the \ion{He}{i} inversions.
       Superimposed black (blue) arrows indicate the Helium (Silicon)
       horizontal fields in the local frame of reference. The white contour along the
       $y$-axis encloses horizontal fields according to the criterion of
       inclinations between $75^\circ < \gamma_\mathrm{He} < 105^\circ$. (c,e)
       Histogram densities of the inferred azimuth differences between Helium
       and Silicon, $\phi_\mathrm{He} - \phi_\mathrm{Si}$, only for the arrows
       which are enclosed within the white contour in the (a) and (b) panels.
       (d) Most common azimuth configurations between Helium  and
       Silicon above $y \sim 8$ and $y \sim 12$, for July 3rd and
       5th respectively.}
       \label{Fig:histograms}%
       \end{figure*}

Figure \ref{Fig:histograms} analyzes the differences between the azimuths inferred 
from the \ion{He}{i} ($\phi_\mathrm{He}$) and \ion{Si}{i} ($\phi_\mathrm{Si}$)
inversions after solving the 180$^\circ$ ambiguity.
Although only two data sets are presented, all of the
other maps show very similar results. Figures
\ref{Fig:histograms}(a,b) compare the photospheric (blue arrows) and
chromospheric (black arrows) horizontal fields for the two days. 
The white contours in the panels delimit the regions with predominantly 
horizontal magnetic fields. The adopted selection criterion
defined ``horizontal fields'' as those having an inclination with respect
to the local frame of reference in the range of
$75^\circ < \gamma_\mathrm{He} < 105^\circ$. Azimuth 
differences\footnote{Azimuths are always positive and measured counterclockwise}, 
$\phi_\mathrm{He} - \phi_\mathrm{Si}$, in between the white contours are presented 
in the histograms of Figs. \ref{Fig:histograms}(c,e), revealing that: 

\begin{enumerate}
\item On July 3rd (i.e., observing mostly the spine region) both
photospheric and chromospheric fields seem to follow each other, differing by
only 5$^\circ$--10$^\circ$, with the \ion{He}{i} arrows being more aligned with the filament
axis (see Fig. \ref{Fig:histograms}(d)); 

\item On July 5th, on the other hand, it was the photospheric field that 
was more aligned with the PIL, and the azimuths showed a larger discrepancy, of 
10$^\circ$--20$^\circ$ on average; 

\item The $\phi_\mathrm{He}$ angles are systematically
larger than the $\phi_\mathrm{Si}$. 
\end{enumerate}

To understand why the azimuth of the chromospheric vector field always pointed in a 
slightly conterclockwise direction with respect to the photospheric one, 
we provide the inset of Fig. \ref{Fig:histograms}(d).

On July 3rd, the chromospheric magnetic field was aligned with the spine, while the photospheric 
one displayed an inverse configuration, thus, having smaller azimuth angles (angles are measured
 from the $x$-axis
and increase counterclockwise). We interpret this as an indication that the field 
lines were wrapped around the filament axis.
On July 5th, the situation was reversed: the filament axis was delineated by the photospheric field 
lines while the chromospheric ones showed a normal polarity configuration, thus mapping a higher 
part of the structure than two days earlier. Again the chromospheric azimuth angles were
larger than the photospheric ones.

\section{Discussion}

In this paper, we present an extensive study of the vector magnetic
field at two different heights in a filament that lies on top of
the polarity inversion line of an active region. This study is mostly based on TIP-II data
in the 10830\,\AA\ spectral region, but also uses other multiwavelength data sources.
Helium intensity core images clearly showed the presence of a thin
filament spine, that we identify with the filament axis, on top of the PIL on
both days. Our observations reveal that this spine is dominated by strong
homogeneous horizontal fields, typically in the range of 400--500\,G in the
chromosphere and 100\,G smaller in the photosphere underneath.  
Vertical gradients of the field strength with
stronger fields in the upper layers, have already been observed \citep[e.g., in
polar crown prominences as shown by][]{leroy83} and modelled in the past
\citep{aulanier03} and have been ascribed to the presence of dips
\citep{anzer69,demoulin89}.  The vector magnetic field in the filament spine 
at chromospheric heights is
highly sheared (i.e., parallel to the PIL), while the photospheric vector field
below the spine shows a uniformly inverse polarity.  Such a configuration
naturally suggests a flux rope topology. This scenario is also favored
by the \textit{TRACE} at $171$\,\AA, which showed the filament with an inverse-S
shape structure. Such a shape is thought to be an indication of the existence of an underlying
flux rope \citep{gibson02}. This configuration of the spine with inverse
orientation in the photosphere and sheared field lines in the chromosphere,
shows no significant change in the observed two-day interval. 
Evidence for the existence of the filament back on July 1st was
presented in the BBSO H$\alpha$ images (see Fig. \ref{Fig:BBfil}).
This readily shows that the process that created the filament occurred well
before our observations. We tentatively identify this previously existing filament 
with the spine region in our observations, and remark on the fact that it showed almost no evolution
during our observing campaign. 

However, the PIL region was not inactive in this time lapse from the 3rd to
the 5th of July. MDI line-of-sight magnetograms show a widening (on July 3rd)
and a narrowing (on July 5th) of the opposite polarities of the AR. The
photospheric longitudinal magnetic field obtained with TIP-II on July 3rd also
indicates a wide separation of both polarities compared to the more compact PILs
observed on July 5th. On this day, the field of view included not only the
spine region, but also the newly appeared orphan penumbrae and pores. This
region has a magnetic topology {\em fundamentally} different from that of the spine
discussed before. The upper panels of Fig. \ref{Fig:vector5jul} show
chromospheric field lines with a normal polarity configuration (above $y \sim
15$\arcsec). The horizontal fields there are stronger than in the spine,
reaching up to 800\,G \citep[see][]{kuckein09}. This normal polarity
configuration suggests the presence of field lines that directly connect
opposite polarities. On the other hand, and most noticeably, in the lower
panels of Fig. \ref{Fig:vector5jul}, {\em the photospheric field lines are
always pointing along the PIL, with a clear sheared configuration}. The 
strengths of these PIL-aligned fields are large, in the range of 1000--1100\,G.
Now, the shear is seen in the photosphere, below the chromospheric 
arching field lines in a normal configuration. If a flux rope above the
photosphere were found in the spine region, a similar flux rope
would be necessarily sitting underneath it, at photospheric heights. Note that a simple
potential arcade model can be excluded here since {\em no sheared field lines are
expected in the photosphere} for such a configuration. In the scenario that we 
suggest, the Helium Stokes profiles would be mapping the top arching field lines
with a normal configuration, while the axis of the rope, with sheared fields, would be located
in the denser photosphere. The field strength of the structure is large enough
to generate pore-like darkenings and penumbral alignments. Moreover, as these
structures become visible following the widening and closing of the PIL region,
we are tempted to suggest that what we are witnessing is the emergence to
the surface of a flux rope structure. Unfortunately, the limited amount of
observations at our disposal, and the prevailing seeing conditions prevent a
study of the evolution similar to that performed by  \citet{okamoto08,
okamoto09}. However, the simultaneous observations in the
chromosphere and the photosphere that we present in this paper indicate 
that a similar process might be at work.  

There is another subtle indication supporting the existence of a lower
lying flux rope in this part of the FOV observed on July 5th. As stated
before, the spine region displays the same
configuration on both days, in agreement with a flux rope located 
at chromospheric heights. However, there is one
difference between the \ion{Si}{i} absorption images on both days 
(see  Fig. \ref{Fig:TIPmaps}).
While on July 3rd the spine is not seen in the Silicon core images, 
it does become obvious in those from July 5th,  
suggesting that, in this case, the axis of the filament sits at lower heights.
It is as if the proximity of the orphan penumbrae
and pore region (interpreted here as a flux rope in the photosphere), forces
the axis of the main filament to move to lower layers.

Recently \citet{mactaggart10} studied the ``sliding doors'' effect through 3D
magnetohydrodynamic (MHD) simulations. They reported that this effect
originates from the emergence and expansion at the photosphere of a flux rope.
In their simulations, the main axis of the flux bundle emerges, due to the
magnetic buoyancy instability, to heights just slightly above the photosphere.
For this area of the observed map (i.e., excluding the spine), we tentatively
propose a scenario where an emerged flux rope is trapped (or slowly evolving)
in the photosphere, as reported in the simulations of \citet{mactaggart10}. In
this scenario, the observed pores and orphan penumbrae are the white-light
counterparts of a flux rope trapped in the photosphere that is indeed part of
the filament. To our knowledge, {\em this is the first observational evidence of a
trapped flux rope in the photosphere}. Along with the
works of \citet{okamoto08,okamoto09} and \citet{lites10}, this work is the
third example of observations of a ``sliding door'' and flux rope emergence
scenario inside an AR filament, with almost the same time scale ($\sim 1$ day).

One could argue that such an emergence process must leave a fingerprint on a
flux history curve of the active region \citep[as proposed 
recently by][]{vargas11}.  No
such indication is seen in Fig. \ref{Fig:Fluxlosses}, neither for the emergence
in the sliding door phase (on July 3rd), nor on July 4-5th when, at some point,
the trapped flux rope emerged and generated the observed pores and orphan penumbrae
system. However, it is important to point out that all these processes largely
involve {\em transverse} fields to which the
longitudinal magnetograms from \textit{SOHO}/MDI are blind. As long as the longitudinal
flux involved in any of these events stays below the flux noise level in Fig.
\ref{Fig:Fluxlosses} ($0.3 \times 10^{21}$\,Mx) it will remain undetectable. This
can be accomplished by flux ropes that are not highly twisted (with a dominant
axial flux over a poloidal component). For example, the case shown by
\citet{vargas11} involves a flux increase of $0.1 \times 10^{21}$\,Mx, which
would be buried in the noise of our flux curves.

The filament formation model by \citet{balle89} and the model of active region
evolution by \citet{vanballe07} form, through successive reconnections above
the photosphere, a twisted flux rope with a strong shear at the axis. The
reconnection gives rise to two types of field lines, sheared field lines that
form the axis of the helical rope, and normally oriented loops below them
\citep[see Figures 1 and 5 in][respectively]{balle89, vanballe07}. In these
models, the photosphere acts as barrier for the submergence of sheared fields,
which must stay in the corona once created by reconnection. However, the loops
with a normal configuration must submerge below the photosphere and cross this
layer at some point. This submergence is driven by tension forces pointing down
and the involved spatial scales are small \citep[900\,km, see Figure 1
in][]{balle89}. While these scales were not easily accessible to quantitative
vector magnetograms in the past, the data presented in this paper achieve the
needed resolution and can be used to search for evidence of these reconnected field
lines. We did not find such a normally oriented component in our photospheric
vector magnetic field maps. This is not only true for the inversions done with
binned data, but also for the inversions done of the 
\ion{Si}{i} line with the original resolution (1 \arcsec). 
Only inverse orientation (in the spine
region) or sheared field lines (in the orphan penumbral region) have been
detected at photospheric heights. In this sense, our observations fail to
support the generation of these normal polarity submerging magnetic field lines. 
Another prediction of the aforementioned models is that velocity measurements must clearly show
downflows in the photosphere close to the PIL, where the field lines submerge.
Therefore, Doppler measurements are definitely needed to verify whether downflows or
upflows of some kind are present at the base of the filaments. This will be
described in Paper II of this series.

Note that the scenario proposed in various recent works
\citep[see, e.g.,][for a review]{archontis08,mackay10}
where the original flux rope remains confined to photospheric
heights and a secondary flux rope is produced after reconnection of the
emerged top part of the original one, could well apply to
our results. The spine region would correspond to the secondary flux rope 
in the chromosphere while the orphan penumbral region would belong
to the parent flux rope that stays deeper down
in the denser layers.What remains unexplained by this model is the 
further disappearance by July 6th of the pores and orphan penumbrae
observed in the \textit{SOHO}/MDI continuum image (Fig. \ref{Fig:MDIcont}).

\section{Conclusions}

The main conclusions of the present work are:

\begin{enumerate}
\item The ``sliding doors'' effect described by Okamoto et al. (2008) 
was seen in our observations during the period between July 3rd and 
July 5th. This has been identified in \textit{SOHO}/MDI data as well 
as in the TIP-II \ion{Si}{i} magnetograms for those days.

\item The observed AR filament can be separated into two areas. The first one
is well observed on July 3rd and shows the filament axis, or spine, in the
Helium absorption images. The \ion{He}{i} data show the magnetic field 
(with horizontal field strengths of 400--500\,G) aligned with the filament, 
indicating a strong sheared configuration. This region is also observed
in the FOV of July 5th, when portions of the spine are even seen in the 
Silicon line core images. This would indicate that the part of the spine 
observed this day lies in lower layers than the portion observed two 
days earlier.

\item The photospheric fields for this area of the filament show an 
inverse configuration of the magnetic field lines that we interpret to 
be the bottom of a flux rope structure. Similarly, we interpret the spine
seen in Helium as the signature of the flux rope axis.

\item The second area corresponds to the orphan penumbrae that 
appear during the sliding door timeframe and clearly shows a different
magnetic configuration. Helium observations exhibit a normal polarity 
configuration whereas the Silicon data show a strongly sheared region
with very intense horizontal fields. This magnetic topology is interpreted 
as a scenario in which the chromospheric magnetic fields trace the top
part of the flux rope, while the spine -or flux rope axis- is still lying
in the photosphere.

\item The observed orphan penumbrae is thus the photospheric 
counterpart of this active region filament. Since the spine is also 
seen in the Silicon line, we emphasize that active region filaments
can have a clear photospheric signature. In particular, it would be 
interesting to find out if all orphan penumbral regions observed 
inside active regions are related to their filaments.

\item The whole emergence process of these orphan penumbrae does not 
leave a clear imprint in the flux history of the active region. This is 
interpreted as due to the flux system being mostly transverse, with a low 
enough twist to not generate identifiable signals in \textit{SOHO}/MDI 
longitudinal magnetograms.

\item We do not observe flux loops with a normal polarity configuration 
in the photosphere, as suggested by some filament generation models based 
on footpoint motions and reconnection \citep[see][]{balle89, vanballe07}. 
While the preexisting filament observed on July 1st in the BBSO data could 
have been created in the way described in these works, the configuration 
and the evolution described here for the period from July 3rd to July 5th
suggests otherwise.

\end{enumerate}

The current study was limited, not only by the small FOV of TIP-II, which
nowadays has a slit twice as long as during our observing campaign in 2005, but
also by the observational gap on July 4th. Magnetic field extrapolations could
undoubtedly shed more light on clarifying the magnetic structure of this
filament. It is now crucial to carry out more multiwavelength measurements, as the one
presented in here, with higher cadence and bigger FOVs to fit the pieces
of the puzzle together, in order to fully understand the origin, evolution and
magnetic topology of AR filaments. In particular, continuous vector magnetograms 
of active regions together with simultaneous imaging of the corona should be able to
prove/disprove the proposed scenario for the last stages of their evolution. The
instrument suite on board the NASA/SDO satellite is the best candidate for such an
study.

\begin{acknowledgements}
We thank an anonymous referee that has greatly enhanced the scientific
content of the paper and its clarity.
This work has been partially funded by the Spanish Ministerio
de Educaci\'on y Ciencia, through Project No. AYA2009-14105-C06-03. Financial support 
by the European Commission through the SOLAIRE Network (MTRN-CT-2006-035484) and help received by C.
Kuckein during his stay at HAO/NCAR are gratefully acknowledged.
Based on observations made with the VTT operated on the island of Tenerife by
the KIS in the Spanish Observatorio del Teide of the Instituto de Astrof\'isica
de Canarias. The authors thank L. Yelles Chaouche, F.
Moreno-Insertis and G. Aulanier for helpful discussions. We also thank B. Lites for 
extensive comments on the manuscript. The Dutch Open
Telescope is operated by Utrecht University at the Spanish Observatorio del
Roque de los Muchachos of the Instituto de Astrof\'isica de Canarias. SOHO is a
project of international cooperation between ESA and NASA. The National Center
for Atmospheric Research (NCAR) is sponsored by the National Science Foundation
(NSF). SOLIS data used here are produced cooperatively by NSF/NSO and NASA/LWS. 
Data from the Big Bear Solar Observatory, New Jersey Institute of Technology, are gratefully acknowledged.
\end{acknowledgements}

\bibliographystyle{aa} 
\bibliography{esquema1} 

\end{document}